\documentclass[a4paper]{article}

\usepackage{icrc2013}
 \usepackage[english]{babel}

\title{Active Galaxies Studied with the H.E.S.S. Observatory}

\shorttitle{Active Galaxies \& H.E.S.S.}

\authors{
\L ukasz Stawarz$^{1,\,2}$, {\it on behalf of the H.E.S.S. Collaboration}
}

\afiliations{
$^1$ Astronomical Observatory, Jagiellonian University, ul. Orla 171, 30-244 Krak\'ow, Poland \\
\scriptsize{
$^{2}$ now at: Institute of Space and Astronautical Science, JAXA, 3-1-1 Yoshinodai, Chuo-ku, Sagamihara, Kanagawa 252-5210, Japan
}
}

\email{stawarz@astro.isas.jaxa.jp}

\abstract{X-ray selected BL Lac objects dominate the population of extragalactic sources detected in the very high energy (VHE; photon energies $\varepsilon_{\gamma} >$\,100GeV) $\gamma$-ray regime with ground-based Cherenkov Telescopes. These are not the only extragalactic VHE emitters, however, since also other types of active galactic nuclei (AGN) have recently been established as VHE sources as well. They include radio galaxies, optically-selected BL Lacs, and Flat-Spectrum Radio Quasars. Here we review the current status of the ongoing AGN studies with the H.E.S.S. instrument, concentrating on the following main three topics: (i) constraints on the structure of the blazar emitting zone, and on the energy dissipation processes involved in the production of the high-energy emission in AGN, enabled by detailed timing and spectral analysis of the H.E.S.S. data; (ii) constraints on the extragalactic background light (EBL) using H.E.S.S. observations of blazars, and (iii) the phenomenological picture of the ``$\gamma$-ray loud'' AGN population emerging from combined VHE and mutli-wavelength observations of different classes of active galaxies.}


\begin{document}
\maketitle

\section{H.E.S.S. \& Active Galaxies}

H.E.S.S. experiment is a system of Imaging Atmospheric Cherenkov Telescopes located in Namibia, and sensitive to $\gamma$-rays with energies from $\simeq 100$\,GeV to $>10$\,TeV. The initial phase of the experiment, consisting of four telescopes working in the stereo mode, is operational since December 2003; the recently completed fifth telescope is operational since July 2012, marking the beginning of the ``H.E.S.S.\,II'' phase. With its superior sensitivity ($\sim 1\%$ of the Crab flux for 25\,h exposure on a point source near zenith), high angular resolution ($\sim 0.1$\,deg for individual photons), as well as excellent energy resolution ($<20\%$), H.E.S.S. is actively pursuing observations of AGN in the Southern hemisphere.

Till now, 24 extragalactic sources have been detected in the VHE range with H.E.S.S., including starburst galaxy NGC 253, flat-spectrum radio quasar PKS 1510--089, two radio galaxies M 87 and Cen A, as well as 20 blazars of the BL Lac type. These are all listed in \cite{cerruti} except of the three recently discovered objects PKS 1440--389 \cite{1440}, PKS 0301--243 \cite{0301}, and KUV 00311--1938 \cite{00311}. Constraining upper limits have been also derived for a number of other AGN of different types \cite{ULs}, including the particular case of Hydra A radio galaxy located in the center of a rich cluster \cite{hydra}. In this short review we summarize the selection of the most recent results presented by the H.E.S.S. Collaboration during the last two years on the VHE-emitting active galaxies.

\section{Recent Results}

\subsection{BL Lacs}

BL Lacs for which the X-ray spectra are dominated by the synchrotron emission of nuclear jets --- the so-called `high-frequency peaked BL Lacs' (HBLs) --- dominate the population of extragalactic sources detected at TeV energies, in agreement with the earlier expectations \cite{costamante}. Recently, the all-sky survey in the high-energy range (HE; $\varepsilon_{\gamma} <100$\,GeV) by the Large Area Telescope LAT onboard the \textit{Fermi Gamma-ray Space Telescope} \cite{LAT}, as well as the operation of a number of modern high-sensitivity X-ray, optical, infrared and radio facilities, enabled for the first truly multi-wavelength (MWL) campaigns targeting VHE-emitting blazars, and not restricted to only the brightest objects during their flaring states. Thus obtained detailed characterization of the broad-band spectra of HBLs, particularly and most notably at $\gamma$-ray frequencies, as well as a much improved view on their MWL variability properties, allowed one to progress on the understanding of the physics of relativistic AGN jets in general.

\begin{figure}[th!]
\centering
\includegraphics[width=\columnwidth]{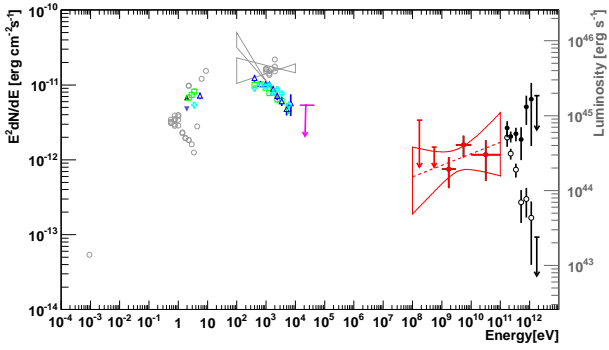}
\caption{Average SED of 1ES 0414+009 from the 2005/09 observations (see \cite{0414}).
H.E.S.S. black filled/open circles with/without EBL correction, respectively.}
\label{f-0414}
\end{figure}

\textit{1ES 0414+009 ($z=0.287$):} The 2005/09 observations of this HBL-type blazar have been presented in \cite{0414}. The source was detected at the significance level of $7.8\,\sigma$ after 74\,h exposure, with the photon index $\Gamma_{\rm VHE} = 3.45 \pm 0.25_{\rm stat} \pm 0.20_{\rm sys}$, and the integrated photon flux $F(\varepsilon_{\gamma}>0.2\,{\rm TeV}) =  1.88 \times 10^{-12}$\,cm$^{-2}$\,s$^{-1}$. No evidence for a variability within the TeV range was found in the collected dataset. The H.E.S.S. detection of this relatively distant object confirmed a low density of the EBL, close to the lower limit from galaxy counts, as otherwise the VHE\,emission of the source would have been attenuated by the photon-photon annihilation below the detection threshold. A very smooth and flat EBL-deabsorbed HE/VHE spectrum of 1ES 0414+009, best characterized as a power-law with photon index $\Gamma_{\gamma} \simeq 1.86$, implies the position of the high-energy peak in the spectral energy distribution (SED; i.e., $\nu-\nu F(\nu)$ representation) above a few TeV (see Figure\,\ref{f-0414}). This, together with the general spectral shape of the optical/X-ray continuum, seriously challenges the standard one-zone synchrotron self-Compton (SSC) emission scenario widely applied in modeling the broad-band spectra of BL Lacs \cite{tavecchio}, as discussed in \cite{0414}.

\textit{1RXS J101015.9--311909 ($z=0.140$):} The 2006/10 observations of this HBL have been presented in \cite{1010}. The source was detected at the significance level of $7.1\,\sigma$ after 49\,h exposure, with $\Gamma_{\rm VHE} = 3.08 \pm 0.42_{\rm stat} \pm 0.20_{\rm sys}$, and $F(\varepsilon_{\gamma}>0.2\,{\rm TeV}) =  2.35 \times 10^{-12}$\,cm$^{-2}$\,s$^{-1}$. No evidence for a variability within the TeV range was found in the collected dataset. The accompanying observations with \textit{Swift} X-ray Telescope revealed some hints for the intrinsic absorption of the soft X-ray continuum of the source by the neutral gas with the hydrogen column number density $N_{\rm H} \simeq 10^{21}$\,cm$^{-2}$. If confirmed, this would constitute a unique finding for the BL Lac class of blazars (but see also \cite{kataoka}). The tentative status of the detection of the absorption features in the X-ray spectrum of 1RXS J101015.9--311909 complicates the modeling of the averaged non-simultaneous SED of the source. Different versions of the applied SSC model fits indicated however in accord that, in the framework of the one-zone approximation, the blazar emission zone has to be only weakly magnetized, with the radiating electrons and magnetic field comoving energy densities ratio of the order of $u'_e/u'_B \simeq 10-100$. This, in fact, is a commonly encountered situation when the averaged broad-band spectra of BL Lacs are modeled in the framework of the one-zone SSC scenario, and similar results have been obtained for the other two HBLs recently studied by the H.E.S.S. collaboration, namely SHBL J001355.9--185406 ($z=0.095$) observed for 41\,h during the period 2008/11 \cite{0013} and PKS 0447--439 ($z <0.59$) detected with H.E.S.S. after the 13\,h exposure in 2009 \cite{0447}.

\begin{figure}[th!]
  \centering
  \includegraphics[width=\columnwidth]{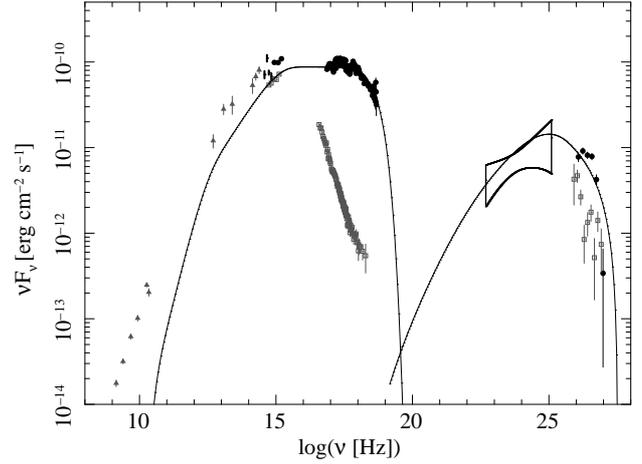}
  \caption{Spectral energy distribution of PKS 2005--489 during the 2009 MWL campaign (filled black symbols), together with the historical data (gray open symbols; see \cite{2005}). The H.E.S.S. spectra are corrected for the EBL absorption.}
  \label{f-2005}
 \end{figure}
 
\textit{PKS 2005--489 ($z=0.071$):} The 2009 observations of this HBL, presented in \cite{2005}, were triggered by the particularly high level (historical maximum) of its synchrotron emission at optical and X-ray frequencies. The source was detected at the significance level of $16\,\sigma$ after 13\,h exposure, with the integrated photon flux $F(\varepsilon_{\gamma}>0.3\,{\rm TeV}) =  6.7 \times 10^{-12}$\,cm$^{-2}$\,s$^{-1}$ and the VHE spectrum best described as a flat power-law continuum $\Gamma_{\rm VHE} = 1.3 \pm 0.6_{\rm stat} \pm 0.2_{\rm sys}$ moderated by the exponential cutoff at $\varepsilon_{\gamma} = 1.3 \pm 0.5$\,TeV. The variability was firmly established in the TeV range on the timescales from days to years. The uniqueness of the source has been revealed by the comparison of the gathered 2009 broad-band dataset with the historical one corresponding to the previous MWL observations in 2004/07. In particular, as shown in Figure\,\ref{f-2005}, the X-ray emission of PKS 2005--489 has changed dramatically during those\,two epochs, both in spectral shape and in the normalization, while only small-amplitude flux changes have been noted in the VHE range. As a result of the exceptionally high X-ray state of the source in 2009, the applied standard one-zone SSC modeling of a quasi-simultaneous SED did not require large departures from the energy equipartition condition, unlike in the majority of flaring BL Lacs. On the other hand, the returned model fit parameters were in general inconsistent with the observed long-term variability properties of the target \cite{2005}.

\begin{figure*}[!t]
\centering
\includegraphics[width=\textwidth]{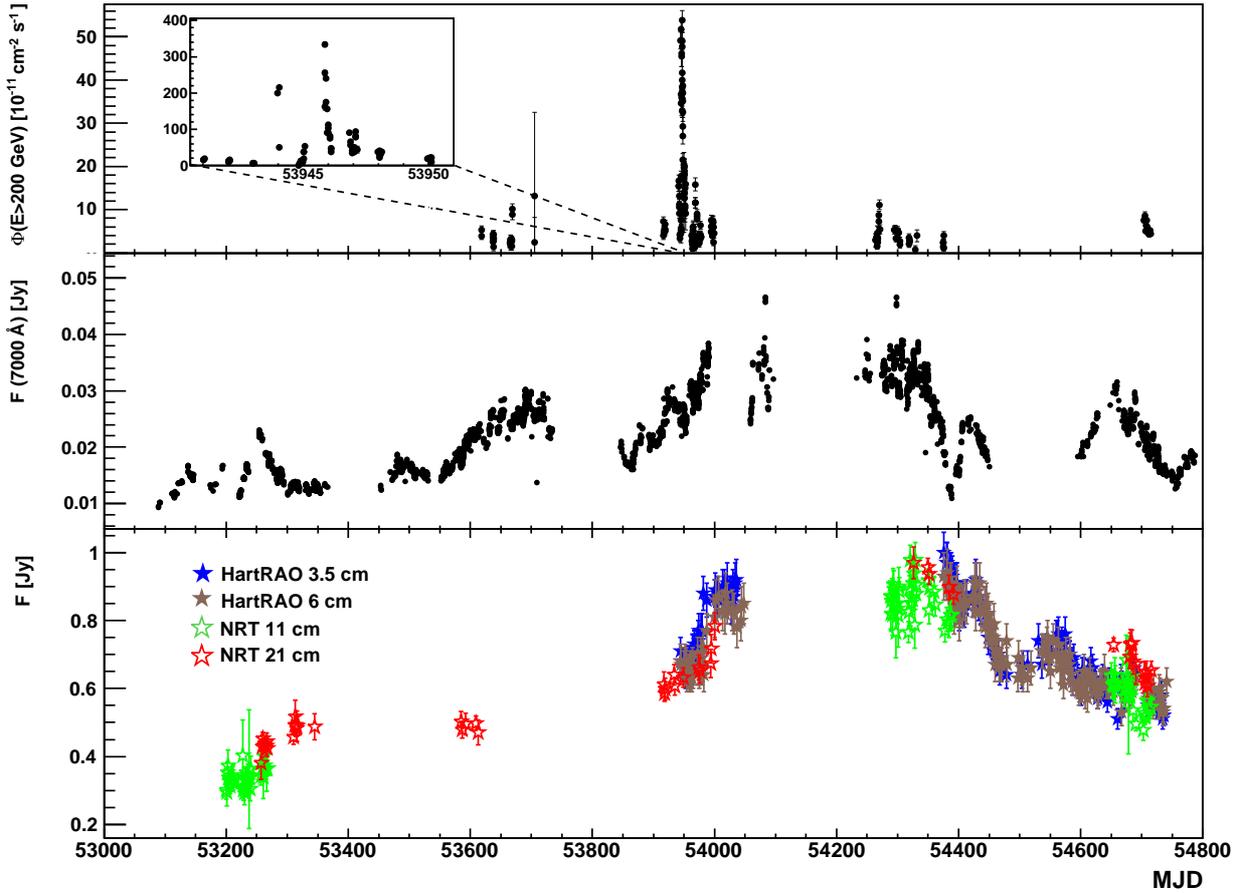}
\caption{Long-term light curves of PKS 2155--304 from VHE $\gamma$-ray measurements (upper panel: H.E.S.S.),  optical data (middle panel: ROTSE) and radio data (lower panel: NRT and HartRAO). See \cite{2155} for details.}
  \label{f-2155}
 \end{figure*}

\textit{PKS 2155--304 ($z=0.116$):} This famous HBL has been monitored with H.E.S.S. multiple times since its first detection in 2002. The source was initially found in its quiescence till the exceptional $\gamma$-ray outburst occurred in July 2006, with the VHE flux 10 times above the average level, and the VHE flux doubling timescales of the order of a few minutes \cite{flare1,flare2}. The MWL data collected around the time of the 2006 flare were recently presented and extensively discussed in \cite{2155} (see also \cite{2155a}). These revealed strong X-ray/VHE flux correlation at high flux levels, weakening however at lower flux levels, and characterized in addition by much smaller X-ray flux changes (factor $\sim 2$) than those observed in the VHE range. No exact optical--VHE correlations were found in the 2006 flaring data, although some relation between the long-term flux evolution at low (optical/radio) frequencies and VHE flaring events was noted (\cite{2155}; see Figure\,\ref{f-2155}). Various approaches to the modeling of the source SED indicated that the one-zone stationary SSC scenario can account for the source spectrum during the source quiescence. Meanwhile, the flaring state which appears physically distinct from the low-activity state, being characterized by very different MWL correlation patterns and spectral evolution, requires more complex emission models, consisting for example of non-homogeneous jet structure and continuous (turbulent) acceleration of the radiating electrons \cite{katarzynski,boutelier}.

All in all, the most recent H.E.S.S. studies of HBL type of blazars reveal distinct MWL variability patterns in different objects (e.g., PKS 2155--304 versus PKS 2005--489), or even during different epochs for the same source, often with no obvious correlation between VHE, HE, X-ray, and optical/UV bands. The EBL-deabsorbed $\gamma$-ray spectra of the targets are found to be also very diverse, being characterized by power-law, exponential cutoff, or smoothly curved distributions. All such diversity, together with the detection of ultra-rapid VHE variability in PKS 2155--304, challenge simple one-zone SSC scenarios widely invoked in the modeling of the broad-band SEDs of HBL type of blazars. It is therefore not clear if the model parameters typically derived in the framework of such modeling --- namely significant departures from the energy equipartition $u'_e/u'_B \geq10$, high minimum Lorentz factors of the radiating electrons $\gamma_e \geq 100$, and emission region linear sizes $\sim (0.3-30) \times 10^{16}$\,cm --- constitute any realistic description of the blazar emission zone in the studied objects.

\subsection{EBL Imprint}

\begin{figure}[th!]
  \centering
  \includegraphics[width=\columnwidth]{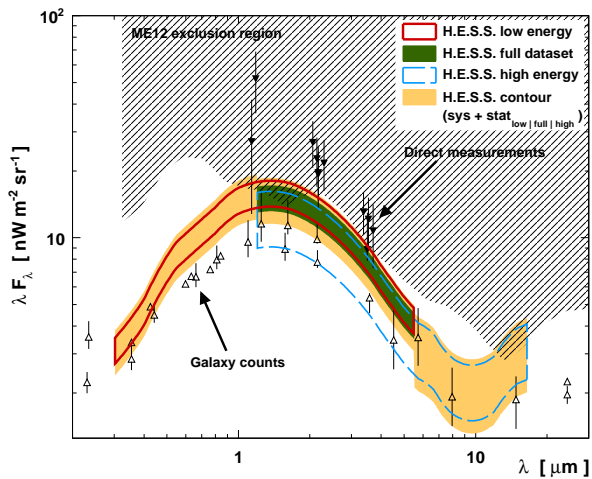}
  \caption{Flux density of the extragalactic background light versus wavelength. The 1$\sigma$ (statistical) contours derived for several energy ranges are described in the top-right legend. The systematic uncertainty is added quadratically to the statistical one to derive the H.E.S.S. contour (see \cite{EBL} for details).}
  \label{f-EBL}
 \end{figure}

As noted above, the detection of distant BL Lacs within the VHE range confirms a low density of the EBL, close to the lower limit from galaxy counts. But the improved statistics of the TeV-detected HBLs, and a precise characterization of their HE and VHE spectra, enables for a more quantitative approach to the problem, i.e. for a precise constraining of the EBL density spectral distribution $n_{\rm EBL}(\varepsilon)$, which is still a subject of an ongoing debate. Such an analysis, consisting of a joint fit of the EBL optical depth $\tau(\varepsilon_{\gamma}, z, n_{\rm EBL})$ and of the intrinsic spectra $\phi_{\rm int}(\varepsilon_{\gamma})$ for H.E.S.S. blazars detected at significance levels $>10\,\sigma$, has been presented in \cite{EBL}. In the analysis, the observed spectra were assumed to be parameterized as $\phi (\varepsilon_{\gamma}) = \phi_{\rm int}^{\alpha}(\varepsilon_{\gamma}) \times \exp\left[ \alpha \times \tau(\varepsilon_{\gamma}, z, n_{\rm EBL}) \right]$, with the EBL template distribution $n_{\rm EBL}(\varepsilon)$ taken from \cite{franceschini}, and the scaling factor $\alpha$ being a free parameter of the model. The intrinsic curvature in the source $\gamma$-ray continua was accounted for by considering various parameterizations of spectral shapes. The likelihood $\mathcal{L}(\alpha)$ of the EBL optical depth normalization and of the intrinsic spectral model was then compared with the hypothesis of a null EBL absorption $\mathcal{L}(\alpha=0)$ by means of the test statistic ${\rm TS} = 2 \times \log \left[\mathcal{L}(\alpha = \alpha_0)/\mathcal{L}(\alpha=0)\right]$. As a result (see Figure\,\ref{f-EBL}), it was found that the EBL template scaled by $\alpha_0 = 1.27^{+0.18}_{-0.15}$ is preferred over a null optical depth $\alpha=0$ at the $\sqrt{\rm TS} \simeq 8.8\,\sigma$ level, and over the un-scaled template $\alpha =1$ at the $\sqrt{\Delta {\rm TS}} \simeq 1.8\,\sigma$ level.

\subsection{Other Types of AGN}

During the last years the H.E.S.S. collaboration presented also results on various types of active galaxies other than BL Lacs. This included upper limits for the VHE\,emission of a luminous radio galaxy Hydra A located at the center of a rich cluster of galaxies \cite{hydra}. These limits, together with the analogous ones in the HE regime, allow to constrain the energetics and the content of the extended lobes formed in the system due to the interactions of relativistic jets with the intracluster medium. In particular, it was found that the extended lobes in Hydra A have to be close to the equipartition condition $u_e/u_B \simeq 1$, and cannot be fully supported by hadronic cosmic rays, as otherwise the derived $\gamma$-ray upper limits would be violated. However, ultrarelativistic protons can still remain a viable pressure support agent to sustain the lobes against the thermal pressure of the ambient medium.

10 years of the MWL observations of the low-power nearby radio galaxy M 87 have been summarized and re-analyzed in the joint H.E.S.S., VERITAS, MAGIC, and \textit{Fermi}-LAT paper \cite{M87}. The details and the relevance of the gathered exceptional dataset for the source is discussed by Mazin et al. \cite{mazin} during this conference.

Finally, the VHE detection of the flat-spectrum radio quasar PKS 1510--089 ($z=0.361$) has been presented in \cite{1510}. The source was detected at the significance level of $9.2\,\sigma$ after 16\,h exposure in 2009, with very steep spectrum $\Gamma_{\rm VHE} = 5.4 \pm 0.7_{\rm stat} \pm 0.3_{\rm sys}$, and $F(\varepsilon_{\gamma}>0.15\,{\rm TeV}) =  1.0 \times 10^{-11}$\,cm$^{-2}$\,s$^{-1}$. The relevance of the discovery follows from the fact that in the case of quasars, which unlike BL Lac objects accrete at high rates and are therefore characterized by a very intense circumnuclear infrared--to--X-ray photon field, VHE $\gamma$-ray photons produced close to the central engine, are in principle expected to be intrinsically absorbed via the photon-photon annihilation (see the discussion in \cite{tanaka}). Therefore, the H.E.S.S. detection of PKS 1510--089, along with the previous MAGIC discoveries of the rapid VHE flares in the two other flat-spectrum radio quasar 3C 279 and PKS 1222+21 \cite{3C279,1222}, indicate that our current view on the location and the structure of the $\gamma$-ray emission zone in luminous blazars may not be accurate. We note that even though the H.E.S.S. data provided insufficient evidence for the VHE variability in PKS 1510--089, exceptionally rapid $\gamma$-ray flux changes in the source have been recently reported in the HE range (doubling timescale $<1$\,h; \cite{saito}).

\section{Conclusions}

The recent H.E.S.S. studies of VHE-emitting active galaxies indicate a need for a MWL approach to understand the physics of blazar sources. The gathered data along with their in-depth analysis reveal at the same time serious limitations of the standard one-zone emission models typically applied for the TeV-detected BL Lacs and quasars. More complex scenarios are therefore needed to account for the observed variety in MWL variability patterns and the broad-band spectra of relativistic jets. Such scenarios can be constrained now much better than before thanks to the VHE detections of (and also upper limits derived for) AGN of different types, including radio galaxies which constitute the parent population of blazar sources. More such studies are expected to follow soon with H.E.S.S.\,II, and this regards also cosmological distant objects, allowing for a further systematic investigation of EBL spectral distribution.

\vspace*{0.5cm}
\footnotesize{{\bf Acknowledgment:}{
The support of the Namibian authorities and of the University of Namibia
in facilitating the construction and operation of H.E.S.S. is gratefully
acknowledged, as is the support by the German Ministry for Education and
Research (BMBF), the Max Planck Society, the German Research Foundation (DFG), 
the French Ministry for Research,
the CNRS-IN2P3 and the Astroparticle Interdisciplinary Programme of the
CNRS, the U.K. Science and Technology Facilities Council (STFC),
the IPNP of the Charles University, the Czech Science Foundation, the Polish 
Ministry of Science and  Higher Education, the South African Department of
Science and Technology and National Research Foundation, and by the
University of Namibia. We appreciate the excellent work of the technical
support staff in Berlin, Durham, Hamburg, Heidelberg, Palaiseau, Paris,
Saclay, and in Namibia in the construction and operation of the
equipment. \\
}}

\end{document}